    \documentstyle[12pt]{article}

    \centerline{\bf  {\Large Non-Abelian (2,0)-Super-Yang-Mills}} 
    \centerline {\bf{\Large Coupled to Nonlinear $\sigma$-Models}} 
    \vspace{0.7cm} 
    \centerline{M. S. G\'oes-Negr\~ao \footnote{
    negrao@gft.ucp.br and negrao@cbpf.br}, J. A. Helay\"el-Neto \footnote{helayel@gft.ucp.br and helayel@cbpf.br
    }and  M. R. Negr\~ao \footnote{ 
    guida@gft.ucp.br and guida@cbpf.br}}  \vspace{.5cm} 

    \centerline{Universidade Cat\'olica de Petr\'opolis(UCP-GFT)}
	\centerline{Centro Brasileiro de Pesquisas F\'{\i}sicas(CBPF-CCP)}

    {
\begin{document} 
     
    \hspace\parindent

    \begin{abstract} 
    \vspace{0.3cm} 
     
    {\footnotesize{\bf  Following a previous work on Abelian (2,0)-gauge theories, one
    reassesses here the task of coupling (2,0) relaxed Yang-Mills super-potentials to a 
    (2,0)-nonlinear $\sigma$-model, by gauging the isotropy or the isometry group of the latter. One pays special
    attention to the extra ``chiral-like" component-field gauge potential that comes out from the relaxation of constraints.}} 
    \end{abstract}

    PACS numbers: 11.10.Lm, 11.15.-q and 11.30.Pb  
     
    \newpage

     In a previous paper \cite{20SYM}, one has investigated the dynamics and the
    couplings
    of a pair of Abelian vector potentials of a class of $(2,0)$-gauge super multiplets     (\cite{hull}-\cite{dine}) whose symmetry lies on a single U(1) group. Since a number
    of     interesting
    features came out, it was a natural question to ask how these fields
    would behave if the non-Abelian version of the theory was to be considered.

    We can see that some subtle changes indeed occur. As we wish to make a full
    comparison between the two aspects(Abelian and non-Abelian) of the same sort of
    theory, all the general set up of the original formulation was kept. The
    coordinates we choose to parametrise the $(2,0)$-superspace are  given by: 
    \begin{eqnarray}  z^A  \equiv (x^{++}, x^{--}; \theta, \bar  \theta), 
    \label{coord}  \end{eqnarray}  where $x^{++}$, $x^{--}$ denote the usual
    light-cone variables,  whereas $\theta$, $\bar \theta$ stand for complex
    right-handed Weyl  spinors. The supersymmetry covariant derivatives are taken
    as:  \begin{eqnarray}  D_{+} \equiv {\partial}_{{\theta}} + i \bar \theta
    {\partial}_{++}  \end{eqnarray}  and 
    \begin{eqnarray} 
    \bar D_{+} \equiv {\partial}_{{\bar \theta}} 
    + i \theta {\partial}_{++}, 
    \label{susycode} 
    \end{eqnarray} 
    where ${\partial}_{++}$ (or ${\partial}_{--}$) represents the derivative 
    with respect to the space-time coordinate $x^{++}$ (or $x^{--}$). 
    They fulfill the algebra: 
    \begin{eqnarray} 
    D_{+}^{2} = {\bar D}_{+}^{2} = 0 \hspace{2.0cm} 
    \{ D_{+},{\bar D}_{+} \} = 2i 
    {\partial}_{++}. 
    \label{dquad} 
    \end{eqnarray} 
    With these definitions for $D$ and ${\bar D}$, one can check that: 
    \begin{eqnarray} 
    e^{i{\theta}{\bar\theta}{{\partial}_{+}}}D_{+}e^{-i{\theta}{\bar\theta}{{\partial}_{+}}} 
    = {{\partial}_{\theta}}, 
    \end{eqnarray} 
    \begin{eqnarray} 
    e^{-i{\theta}{\bar\theta}{{\partial}_{+}}}{\bar 
    D}_{+}e^{i{\theta}{\bar\theta}{{\partial}_{+}}} = 
    {{\partial}_{\bar\theta}}. 
    \end{eqnarray} 
     
    The fundamental non-Abelian matter superfields are the 
    scalar and left-handed spinor superfields, both subject to the chirality constraint; their respective 
    component-field expressions are given by: 
    \begin{eqnarray} 
    {\Phi}^{i}(x;\theta,\bar \theta)&=& 
    e^{i\theta{\bar\theta}{\partial}_{++}}(\phi^{i} +\theta 
    \lambda^{i}),\nonumber\\ {\Psi}^{I}(x;\theta,\bar \theta)&=& 
    e^{i\theta{\bar\theta}{\partial}_{++}}(\psi^{I} +\theta \sigma^{I});
    \label{c5} 
    \end{eqnarray} 
    the
    fields $\phi^{i}$ and $\sigma^{I}$ are scalars, whereas
    $\lambda^{i}$ and $\psi^{I}$  stand respectively for
    right- and left-handed Weyl spinors. The indices ${i}$ and ${I}$ label the
    representations where the corresponding matter fields are set to transform under the
        Yang-Mills group. 
     
    We present below the gauge transformations for both $\Phi$
    and $\Psi$, assuming that we are dealing with a {\underline {compact}} and
    {\underline     {simple}} gauge
    group, ${\cal G}$, with generators, ${G_{a}}$, that fulfill the algebra
    $[G_{a},G_{b}]$=$if_{abc}G_{c}$. The transformations read as below:   
    \begin{eqnarray} 
    {\Phi}'^{i}=R(\Lambda)^{i}_{j}\Phi^{j}, \hspace{1cm}
    {\Psi}'^{I}=S(\Lambda)^{I}_{J}\Psi^{J},  \label{c8}  \end{eqnarray}  where $R$
    and $S$ are matrices that respectively represent a gauge  group element in the
    representations under which $\Phi$ and $\Psi$  transform. Taking into account
    the chiral constraints on $\Phi$ and $\Psi$,  and bearing in mind the
    exponential representation for $R$ and $S$ in terms of the group generators, 
    we find that the gauge parameter superfields, $\Lambda^{a}$, must satisfy the 
    same sort of constraint, namely, they are chiral scalar superfields. They can
    therefore be     expanded as follows: 
    \begin{eqnarray}  
    {\Lambda}^{a}(x;\theta,\bar \theta)&=& 
    e^{i\theta{\bar\theta}{\partial}_{++}}(\alpha^{a} +\theta \beta^{a}), 
    \label{c9}  \end{eqnarray} 
    where $\alpha^{a}$ are scalars and $\beta^{a}$ are right-handed spinors. 
     
    The kinetic action for ${\Phi^{i}}$ and ${\Psi^{I}}$ can be made invariant 
    under the local transformations (\ref{c8}) by minimally coupling the 
    gauge potential superfields, $\Gamma^{a}_{--}(x;\theta,{\bar \theta})$ 
    and $V^{a}(x;\theta,{\bar \theta})$, according to the minimal coupling 
    prescriptions: 
    \begin{eqnarray} 
    S_{inv}=\int d^2 x d\theta d{\bar \theta} \{i[{\bar \Phi}e^{hV} 
    (\nabla_{--} \Phi)- ({\bar\nabla}_{--} \bar\Phi)e^{hV}\Phi]+{\bar\Psi}e^{hV}\Psi\}, 
    \label{c10} 
    \end{eqnarray} 
    where the gauge-covariant derivatives are defined in the sequel. 
     
    The Yang-Mills supermultiplets are introduced by means of the 
    gauge-covariant derivatives which, according to the discussion of ref. 
    \cite{chair}, can be expressed as below: 
    \begin{eqnarray} 
    {\nabla}_{+} &\equiv & D_{+} + {\Gamma}_{+},\hspace{4.0cm} {\bar 
    \nabla}_{+} \equiv {\bar D}_{+}, 
    \\ 
    {\nabla}_{++} &\equiv & {\partial}_{++} + {\Gamma}_{++} 
    \hspace{1.5cm} and \hspace{1.5cm} {\nabla}_{--} \equiv 
    {\partial}_{--} -ig{\Gamma}_{--}, 
    \label{dercov} 
    \end{eqnarray} 
    with the gauge superconnections ${\Gamma}_{+}$, ${\Gamma}_{++}$ and 
    ${\Gamma}_{--}$ being all Lie-algebra-valued. Note that ${\Gamma}_{++}$ does
    not enter the Lagrangian density of eq.(\ref{c10}). The gauge couplings, $g$ 
    and $h$, can in principle be taken different; nevertheless, this would
    \underline{not}  mean that we are gauging two independent symmetries. There is
    a single simple gauge group,  $\cal{G}$, with just one gauge-superfield
    parameter, ${\Lambda}$. It is the particular form of the  $(2,0)$-minimal
    coupling (realised by the exponentiation of ${V}$ and the connection present 
    in ${\nabla}_{--}$) that opens up the freedom to associate different  coupling
    parameters     to the gauge superfields $V$ and
    ${\Gamma}_{--}$. The superpotentials ${\Gamma}_{+}$ and  ${\Gamma}_{++}$ can be both
        expressed in
    terms of the real scalar  superfield, $V(x;\theta,{\bar \theta})$, according
    to \cite{20SYM}:  
    \begin{eqnarray}  {\Gamma}_{+} = e^{-gV}(D_{+} e^{gV}) 
    \label{gamamais}  \end{eqnarray} 
    and 
    \begin{eqnarray} 
    {\Gamma}_{++} 
    = - \frac{i}{2} {\bar D}_{+} [e^{-gV}(D_{+} 
    e^{gV})]. 
    \label{gama++} 
    \end{eqnarray} 
     
    To establish contact with a component-field formulation and to 
    actually identify the presence of an additional gauge potential, we 
    write down the $\theta$-expansions for $V^{a}$ and ${\Gamma}^{a}_{--}$: 
    \begin{eqnarray} 
    V^{a}(x;\theta,{\bar \theta}) = C^{a} + \theta 
    \xi^{a} - {\bar \theta}{\bar \xi^{a}} + \theta{\bar 
    \theta}v^{a}_{++} 
    \label{ve} 
    \end{eqnarray} 
    and 
    \begin{eqnarray} 
    {\Gamma}^{a}_{--}(x;\theta,{\bar 
    \theta}) &=& \frac{1}{2}(A^{a}_{--} + iB^{a}_{--}) + i\theta (\rho^{a} +
    i\eta^{a}) \nonumber\\ &+& i{\bar \theta}  ({\chi}^{a} + i\omega^{a}) +
    \frac{1}{2}\theta{\bar \theta}(M^{a}+iN^{a}).  
    \label{gama--}  
    \end{eqnarray} 
     
    $A^{a}_{--}$, $B^{a}_{--}$ and $v^{a}_{++}$ are the light-cone components of
    the  gauge potential fields; $\rho^{a}, \eta^{a}, {\chi}^{a}$ and $\omega^{a}$
    are  left-handed Majorana spinors; $M^{a}, N^{a}$ and $C^{a}$ are real scalars
    and  $\xi^{a}$ is a complex right-handed spinor. 

    The infinitesimal gauge transfomations for $V^{a}$ and ${\Gamma}^{a}$ are given
    by \begin{eqnarray}
    {\delta}V^{a} = \frac{i}{h}{({\bar\Lambda} - {\Lambda})}^{a} -
    \frac{1}{2}f^{abc}({\bar\Lambda} + {\Lambda})_{b}V_{c}
    \label{dv}
    \end{eqnarray}
    and
    \begin{eqnarray}
    {\delta}{\Gamma}_{--}^{a} = - f^{abc}{\Lambda}_{b}{\Gamma}_{c--} +
    \frac{1}{g}{\partial}_{--}{\Lambda}_{a}.
    \label{dg}
    \end{eqnarray}

    No derivative acts on the ${\Lambda}^{a}$'s in eq.(\ref{dv}); this suggests
    the possibility of choosing a Wess-Zumino gauge for $V^{a}$. If such a
    choice is adopted, it can be shown that the gauge transformations of the
    $\theta$-component fields above read  as shown:  
    \begin{eqnarray}  
    \delta C &=&\frac{2}{h} {\Im m}\alpha, \nonumber\\  
    \delta \xi &=& -\frac{i}{h}\beta,
    \nonumber\\  
    \delta v^{a}_{++}&=&\frac{2}{h} {\partial}_{++} \alpha^{a} -
    f_{abc}\alpha^{b}v^{c}_{++}, \nonumber\\   
    \delta A^{a}_{--}&=&\frac{2}{g}
    {\partial}_{--} \alpha^{a} - f_{abc}\alpha^{b}A^{c}_{--}, , \nonumber\\ 
    \delta B^{a}_{--}&=& - f_{abc}\alpha^{b}B^{c}_{--}, \nonumber\\  
    \delta \eta^{a} &=& - f_{abc}\alpha^{b}\eta^{c}, \nonumber\\  \delta
    \rho^{a} &=& - f_{abc}\alpha^{b}\rho^{c}, \nonumber\\  
    \delta M^{a} &=& - f_{abc}\alpha^{b}M^{c} +
    f_{abc}{\partial_{++}\alpha}^{b}B^{c}_{--},     \nonumber\\  
    \delta N^{a} &=& \frac{2}{g} {\partial}_{++}{\partial}_{--} \alpha^{a} -
    f_{abc}\alpha^{b}N^{c} - f_{abc}{\partial_{++}\alpha}^{b}A_{--}^{c}, 
    \nonumber\\  
    \delta {\chi}^{a} &=& -f_{abc}\alpha^{b}{\chi}^{c},  \nonumber\\ 
    \delta \omega^{a} &=& -f_{abc}\alpha^{b}\omega^{c}.
      \label{t}  
      \end{eqnarray} 
    This gauge variations suggest that we should take $h=g$, so that the 
    $v^{a}_{++}$-component 
    could be identified as the light-cone partner of $A^{a}_{--}$, 
    \begin{eqnarray} 
    v^{a}_{++} 
    \equiv A^{a}_{++}; 
    \label{ve++} 
    \end{eqnarray} 
    this procedure yields a pair of component-field gauge potentials: $A^{\mu} 
    \equiv (A^{0}, A^{1})=(A^{++};A^{--})$ and $B_{--}$; the latter without the $B_{++}$ partner.  
     
    It is interesting to point out that at this stage the first remarkable difference
    between     the
    Abelian and the non-Abelian versions of the theory arises. In the Abelian
    case \cite{20SYM}, it was shown that both fields ${\chi}$ and $\omega$ were
    gauge invariant and the fields $M$ and $N$ could be identified with a
    combination of $A_{--}$ and $B_{--}$. This combination, which was naturally
    dictated by the form of the gauge transformations, ensured the symmetry
    of the Lagrangian. In the present situation, the gauge transformations do not
    undertake that we may express $M$ and $N$ in terms of 
    $A_{--}$ and $B_{--}$, as it was done before. On the other hand, the
    ${\chi}$-and $\omega$-fields are no longer auxiliary fields, contrary to what happens
    in the
    Abelian version.

    To discuss the field-strength superfields, we start by analysing the algebra of
    the gauge covariant derivatives. The former are defined such that:
    \begin{eqnarray} \{{\nabla}_{+},{\nabla}_{+}\} &\equiv& {\cal F} =
    2D_{+}{\Gamma_{+}}, \nonumber\\ 
    \{{\nabla}_{+},{\bar\nabla}_{+}\} &\equiv&
    2i{\nabla}_{++}, \nonumber\\ 
    {[{\nabla}_{+},{\nabla}_{++}]} &\equiv& W_{-} =
    D_{+}{\Gamma}_{++} - {\partial}_{++}{\Gamma}_{+}, \nonumber\\
    {[{\nabla}_{+},{\nabla}_{--}]} &\equiv& W_{+} = -igD_{+}{\Gamma}_{--} -
    {\partial}_{++}{\Gamma}_{+} -ig[{\Gamma}_{+},{\Gamma}_{--}], \nonumber\\
    {[{\bar\nabla}_{+},{\nabla}_{++}]} &\equiv& U_{+},\nonumber\\
    {[{\bar\nabla}_{+},{\nabla}_{--}]} &\equiv& U_{-} = -ig{\bar
    D}_{+}{\Gamma}_{--}, \nonumber\\ {[{\nabla}_{++},{\nabla}_{--}]} &\equiv&
    {\cal Z}_{+-} = -ig{\partial}_{++}{\Gamma}_{--} - {\partial}_{--}{\Gamma_{++}}
    - ig [{\Gamma_{+}},{\Gamma_{--}}]. \label{fs}  \end{eqnarray}

    The results obtained for the field-strengths are consistent with the Bianchi
    identities. The identity for $U_{+}$,
    \begin{eqnarray}
    [ {\bar \nabla}_{+}, \{ {\nabla}_{+},{\bar \nabla}_{+}\} ] + [ {\nabla}_{+},
    \{ {\bar \nabla}_{+},{\bar \nabla}_{+}\} ] + [ {\bar \nabla}_{+}, \{
    {\bar \nabla}_{+},{\nabla}_{+}\} ]=0
    \end{eqnarray}
    gives immediately that $U_{+} = 0$. The Bianchi identity for $Z_{+-}$,
    \begin{eqnarray}
    [ {\nabla}_{--}, \{ {\nabla}_{+},{\bar \nabla}_{+}\} ]+ \{ {\nabla}_{+}, [
    {\bar \nabla}_{+},{\nabla}_{--}] \}  - \{ {\bar \nabla}_{+}, [
    {\nabla}_{--},{\bar \nabla}_{+}] \} = 0,
    \end{eqnarray}
    allows us to express $Z_{+-}$ as
    \begin{eqnarray}
    Z_{+-} = -\frac{i}{2}{\nabla}_{+}U_{-} - \frac{i}{2}{\bar \nabla}_{+}W_{-};
    \end{eqnarray}
    and, finally, the Bianchi identity
    \begin{eqnarray}
    [ {\bar \nabla}_{+}, \{ {\nabla}_{+},{\nabla}_{+}\} ] + [
    {\nabla}_{+}, \{ {\nabla}_{+},{\bar \nabla}_{+}\} ] + [ {\nabla}_{+}, \{ {\bar
    \nabla}_{+},{\nabla}_{+}\} ] = 0
    \end{eqnarray}
    leads to 
    \begin{eqnarray}
    W_{+} = \frac{i}{4} {\bar D}_{+} {\cal F}.
    \end{eqnarray}
    These are the relevant results yielded by pursuing an investigation of the Bianchi
        identities.

    The gauge field, $A_{\mu}$, has its field
    strength, $F_{\mu\nu}$, located at the ${\theta}$-component of the combination
    ${\Omega}{\equiv} W_{-}+{\bar U}_{-}$. This suggests the following kinetic
    action for the Yang-Mills sector: \begin{eqnarray}
    S_{YM} &=&
    \frac{1}{8g^{2}}{\int}d^{2}xd{\theta}d{\bar\theta}Tr{\Omega{\bar\Omega}}
    \nonumber\\
    &=& {\int}d^{2}xTr[\frac{-1}{4}F_{\mu\nu}F^{\mu\nu} +
    \frac{i}{4}{\Sigma}\stackrel{\mathrm{\leftrightarrow}}{\partial}_{++}{\bar\Sigma} +
    \frac{1}{4}M^{2}],  
    \end{eqnarray}  
    where ${\Sigma={\rho +i\eta} + {\bar{\chi}}
    -i{\bar\omega}}$ and
    $A\stackrel{\mathrm{\leftrightarrow}}{{\partial}}B\equiv(\partial 
    A)B-A(\partial B)$.

    Choosing now a supersymmetry-covariant gauge-fixing, instead of the
    Wess-Zumino, we propose the following gauge-fixing term in superspace:
    \begin{eqnarray} S_{gf} &=&
    -\frac{1}{2\alpha}{\int}d^{2}xd{\theta}d{\bar\theta}Tr[{\Pi}{\bar\Pi}]
    \nonumber\\
    &=& -\frac{1}{2\alpha}{\int}d^{2}x\{[({\partial}_{\mu}A^{\mu})^{2} +
    ({\partial}_{\mu}A^{\mu})N + \frac{1}{4}N^{2}] \nonumber\\
    &+& \frac{1}{4}[M^{2} -
    2M{\partial}_{++}B_{--} + ({\partial}_{++}B_{--})^{2}] \nonumber\\ 
    &-&i({\rho + 
    i\eta})\stackrel{\mathrm{\leftrightarrow}}{\partial}_{++}({\bar\rho
    -i\bar \eta})\}, \label{gf}
    \end{eqnarray}
    where ${\Pi}=-iD_{+}{\Gamma}_{--} + \frac{1}{2}D_{+}{\partial}_{--}V$.

    So, the total action reads as follows:
    \begin{eqnarray}
    S &=& {\int}d^{2}xTr\{-\frac{1}{4}F_{\mu\nu}F^{\mu\nu}
    -\frac{1}{2\alpha}({\partial}_{\mu}A^{\mu})^{2} -\frac{1}{2\alpha}
    ({\partial}_{\mu}A^{\mu})N - \frac{1}{8\alpha}N^{2}\nonumber\\ &+&
    \frac{1}{4}(1 - \frac{1}{2\alpha})M^{2} +
    \frac{1}{4\alpha}M({\partial}_{++}B_{--}) - \frac{1}{8\alpha}
    ({\partial}_{++}B_{--})^{2} \nonumber\\  
    &+&\frac{i}{2\alpha}({\rho + 
    i\eta})\stackrel{\mathrm{\leftrightarrow}}{\partial}_{++}({\bar\rho -i\bar
    \eta}) +
    \frac{i}{4}{\Sigma}\stackrel{\mathrm{\leftrightarrow}}{\partial}_{++}{\bar\Sigma}\}. 
        \label{total} 
    \end{eqnarray}

    Using  
    eq.(\ref{total}), we are ready to write down the propagators for $A^{a}$,
    $B^{a}_{--}$, $N^{a}$, $M^{a}$, ${\rho}^{a}$, ${\eta}^{a}$, ${{\chi}}^{a}$ and
    ${\omega}^{a}$:  \begin{eqnarray}  
    \langle AA \rangle &=& -\frac{i}{2\Box(1-\Box)}
    \omega^{\mu\nu},\nonumber\\ 
    \langle BB \rangle &=&
    -\frac{i(2{\alpha}-1)}{4{\alpha}(1-{\alpha})}\frac{{\partial}^{2}_{--}}{{\Box}^2},
    \nonumber\\
    \langle AN
    \rangle &=& \frac{i{\alpha}}{{\Box}(1-{\Box})}
    (1-{\Box}+{\alpha}){\partial}^{\mu},\nonumber\\   
    \langle N{A} \rangle &=&
    \frac{i}{(1-{\Box}){\Box}} {\partial}^{\nu}\nonumber\\
    \langle NN \rangle &=& -\frac{2i{\alpha}^{2}}{(1-{\Box})}\nonumber\\
    \langle MM \rangle &=& -\frac{i}{16{\alpha}}\frac{1}{(1-{\alpha})}\nonumber\\
    \langle MB \rangle &=& -\langle BM \rangle =
    \frac{i}{8{\alpha}(1-{\alpha})}\frac{\partial_{--}}{\Box}\nonumber\\
    \langle (\rho + i\eta)(\bar\rho -i\bar\eta) \rangle &=&
    -\frac{2\alpha}{({\alpha}-1)}\frac{\stackrel{\mathrm{\leftrightarrow}}{\partial}_{--}
    }{4
        {\Box}}\nonumber\\
    \langle (\rho + i\eta)({\chi} +i\omega) \rangle &=&
    -\frac{\alpha}{4}\frac{\stackrel{\mathrm{\leftrightarrow}}{\partial}_{--}}{\Box}     \nonumber\\
    \langle (\bar{\chi} - i\bar\omega)(\bar\rho -i\bar\eta) \rangle &=&
    +\frac{\alpha}{4({\alpha}-1)}\frac{\stackrel{\mathrm{\leftrightarrow}}{\partial}_{--}
    }
    {\Box}\nonumber\\
    \langle (\bar{\chi} - i\bar\omega)({\chi} +i\omega) \rangle &=&
    +\frac{(\alpha+2)}{4}\frac{\stackrel{\mathrm{\leftrightarrow}}{\partial}_{--}}{\Box}.
    \label{prop}  
    \end{eqnarray} 

    Expressing the action of equation (\ref{c10}) in terms of 
    component fields, and coming back to the $(2,0)$-version of the
    Wess-Zumino gauge, the  matter-gauge sector Lagrangian reads:  
    \begin{eqnarray} 
    {\cal L}_{matter-gauge}&=& 2 \phi^{\ast i} \Box {\phi}_{i} -ig
    [{\phi}^{\ast i} A_{--}^{a}(G_a)_{i}^{j} {\partial}_{++}{\phi}_{j}-c.c]+
    {\bar{\sigma}}^{i}{\sigma}_{i} +
    \nonumber\\
    &-& ig [{\phi}^{\ast i} A_{++}^{a} (G_a)_{i}^{j}
    {\partial}_{--}{\phi}_{j}-c.c]- g
    {\phi}^{\ast i}M^{a}(G_{a})_{i}^{j} {\phi}_{j} + \nonumber\\ 
    &-&  \frac{i}{2}g^{2}
    {\phi}^{\ast i} A_{++}^{a} A_{--}^{b}{\phi}_{i} d_{abc}G_{c}- g{\bar 
    \lambda}^{i}A_{--}^{a}(G_a)_{i}^{j}{\lambda}_{j}+\nonumber\\
    &-&\frac{1}{2} {\phi}^{\ast i} A_{++}^{a}B_{--}^{b}{\phi}_{i}f_{abc}
    G_{c} + 2i{\bar\lambda}^{i}
    {\partial}_{--}{\lambda}_{i} + \nonumber\\ &-&ig {\phi}^{\ast
    i}[({{\chi}}^{a} +{\bar \rho}^{a} +i{\omega}^{a} - i{\bar 
    \eta}^{a})(G_a)_{i}^{j}{\lambda}_j  -c.c]
    +\nonumber\\&-&2i{\bar\psi}^{i}{\partial}_{++}{\psi}_i -g {\bar \psi}^{i}
    A_{++}^{a}(G_a)_{i}^{j}{\psi}_{j},  
    \end{eqnarray}
    where ${d_{abc}}$ are the(representation-dependent) symmetric coefficients
    associated to $\{{G}_{a},{G}_{b}\}$.    

    One immediately checks that
    the extra gauge field, $B_{--}$, does  \underline{not}  decouple from the
    matter sector. Our point of view of keeping the 
    superconnection ${\Gamma}_{--}$ as a complex superfield naturally  introduced
    this extra gauge potential in addition to the  usual gauge field, $A_{\mu}$:
    $B_{--}$ behaves as a second gauge field.  The fact that it yelds a massless
    pole of order two in the spectrum may harm unitarity. However, the mixing
    with the $M$-component of ${\Gamma}_{--}$, which  is a compensating field,
    indicates that we should couple them to  external currents and analyse the
    imaginary part of the  current-current amplitude at the pole. In so doing, this
    imaginary  part turns out to be positive-definite, and so no ghosts are 
    present. It is very interesting to point out that, in the Abelian case,
    $B_{--}$ showed the same behaviour \cite{20SYM}. It coupled to $C$ instead of
    $M$, but these two fields show the same kind of behaviour: $C$ (in the Abelian
    case) and $M$ (in the non-Abelian case) are both compensating fields. This
    ensures us to state that $B_{--}$ behaves as a  physical gauge field: it has
    dynamics and couples both to matter and the gauge field $A^{\mu}$. Its  only
    peculiarity regards the presence of a single component in the  light-cone
    coordinates. The $B$-field plays rather the r\^ole of a  ``chiral gauge
    potential''. Despite the presence of the pair of  gauge fields, a
    gauge-invariant mass term cannot be introduced, since  $B$ does not carry the
    $B_{++}$-component, contrary to what  happens with $A^{\mu}$. 

    Let us now turn
    to the coupling of the two  gauge potentials, $A_{\mu}$ and $B_{--}$, to a
    nonlinear  $\sigma$-model, always keeping a supersymmetric scenario. It is
    our main purpose henceforth to carry out the coupling of a  $(2,0)$
    $\sigma$-model to the relaxed gauge superfields of the ref.  \cite{chair}, and
    show that the extra vector degrees of freedom do not  decouple from the matter
    fields (that is, the target space 
    coordinates)\cite{bagger}\cite{kar}\cite{brooks}\cite{hel}. The extra gauge
    potential, $B_{--}$, obtained upon relaxing  constraints can therefore acquire
    a dynamical significance by means  of the coupling between the $\sigma$-model
    and the Yang-Mills fields  of ref.\cite{chair}. To perform the coupling of the
    $\sigma$-model to the Yang-Mills fields we reason along the same
    considerations as i ref.\cite{20SYM} and find out that: \begin{eqnarray} 
    {\cal L}_{\xi} &=& {\partial}_{i}[K(\Phi,\tilde\Phi) - {\xi}(\Phi)  -
    {\tilde\xi}(\tilde\Phi)]{\nabla}_{--}{\Phi}^{i} + \nonumber\\ &-& 
    {\tilde\partial}_{i}[K(\Phi,\tilde\Phi) - {\xi}(\Phi) - 
    {\tilde\xi}(\tilde\Phi)]{\nabla}_{--}{\tilde\Phi}^{i},  \label{calL} 
    \end{eqnarray}  where ${\xi(\Phi)}$ and ${\bar\xi}(\bar\Phi)$ are a pair of
    {\it chiral} and {\it  antichiral} superfields, ${\tilde  \Phi}_{i} \equiv
    exp(i{\bf L}_{V.\bar k}){\bar  \Phi}_{i}$ and ${\nabla}_{--}{\Phi}^{i}$
    and  ${\nabla}_{--}{\tilde \Phi}^{i}$ are defined in perfect analogy to 
    what is done in the case of the bosonic ${\sigma}$-model: 
    \begin{eqnarray} 
    {\nabla}_{--}{\Phi}_{i} \equiv {\partial}_{--}{\Phi}_{i} - g 
    {\Gamma}_{--}^{\alpha} {k}_{\alpha}^{i}(\Phi) 
    \end{eqnarray} 
    and 
    \begin{eqnarray} 
    {\nabla}_{--}{\tilde \Phi}_{i} 
    \equiv {\partial}_{--}{\tilde \Phi}_{i} - g {\Gamma}_{--}^{\alpha} {\bar 
    k}_{\alpha i} (\tilde \Phi). 
    \end{eqnarray} 

    The interesting point we would like to stress is that the extra 
    gauge degrees of freedom accommodated in the component-field 
    $B_{--}(x)$ of the superconnection ${\Gamma}_{--}$ behave as a 
    genuine gauge field that shares with $A^{\mu}$ the feature of coupling to 
    matter and to $\sigma$-model \cite{chair}. This result can be 
    explicitly read off from the component-field Lagrangian projected 
    out from the superfield Lagrangian ${\cal L}_{\xi}$. 
	
	We therefore 
    conclude that our less constrained $(2,0)$-gauge theory yields a 
    pair of gauge potentials that naturally transform under the action 
    of a single compact and simple gauge group and may be consistently coupled to
    matter fields as well as to the $(2,0)$ nonlinear $\sigma$-models by
    means of the gauging of their isotropy and isometry groups. 
	
	Relaxing
    constraints in the $N=1$- and $N=2-D=3$ supersymmetric algebra of covariant
    derivatives may lead to a number of peculiar features of the gauged
    $O(3)$-$\sigma$-model \cite{tripa} in the presence of Born-Infeld terms for the
    pair of gauge potentials which share the same symmetry group; of special interest are
    the     self-dual
    equations that may stem from this model \cite{eu}.

    The authors would
    like to thank M. A. De  Andrade, A. L. M. A. Nogueira and O. Del Cima for
    enlightening discussions; CNPq and Capes are acknowledged for the invaluable
    financial support.

    \end{document}